\documentclass{ws-ijmpd}
\usepackage[super,compress]{cite}
\usepackage{xcolor}
\begin{document}

\markboth{Sut-Ieng Tam}
{Cosmic Structures in CDM and SIDM }

\catchline{}{}{}{}{}
%
%%%%%%%%%%%%%%%%%%%%%%%%%%%%%%%%%%%%%%%%%%%%%%%%%%%%%%%%%%%%%%%%%%%%

\title{Cosmic Structures in CDM and SIDM }

\author{Sut-Ieng Tam}

\address{
Institute of Physics, National Yang Ming Chiao Tung University, No. 1001, Daxue Rd. East Dist., Hsinchu City 300093, Taiwan\\
sitam@nycu.edu.tw}

\maketitle

\begin{abstract}
The standard $\Lambda$ Cold Dark Matter ($\Lambda$CDM) model has achieved remarkable success in explaining the formation and evolution of cosmic structures on large scales, supported by a wide range of observations, including the cosmic microwave background, large-scale structure surveys, and galaxy clusters. However, discrepancies between theoretical predictions and observations on small scales, have motivated the exploration of alternative dark matter models, including the self-interacting dark matter (SIDM) scenario. This review provides an overview of the theoretical foundations of CDM structure formation, the small-scale challenges, and the solutions proposed within the SIDM framework. We summarize recent theoretical developments in the SIDM framework and discuss current observational constraints on the dark matter self-interaction cross-section with particular emphasis on galaxy clusters. Finally, we highlight future prospects for improving our understanding of the fundamental properties of dark matter.
\end{abstract}


\ccode{PACS numbers:}

%\tableofcontents
\section{Introduction}
Dark matter, as the mysterious matter component in our Universe, arose from early astrophysical observations that could not be explained by visible matter alone. Back in the 1930s, Fritz Zwicky \cite{Zwicky} analyzed the velocity dispersion of galaxies in the Coma cluster and found it to be remarkably high, around $\sim 1500\,\mathrm{km/s}$. This velocity implied a total gravitational mass far exceeding that inferred from the luminous matter alone. To account for the discrepancy, Zwicky proposed the existence of an unseen mass component that provides the additional gravitational potential necessary to bind the cluster. Later, Rubin et al. \cite{Rubin} provided another compelling evidence for dark matter through through their measurements of galaxy rotation curves. These curves remained flat at large radii, contrary to the expectation from Newtonian dynamics ($v \propto r^{-1/2}$) if only baryonic matter were present. This suggested the presence of an extended, invisible mass distribution beyond the optical edge of galaxies. Further support came from gravitational lensing observations \cite{Lynds, Soucail} and X-ray measurements \cite{Lea, Cavaliere} of galaxy clusters, both  demonstrated significant mass components beyond the visible matter. 

The Cold Dark Matter paradigm emerged in the early 1980s as a compelling framework to explain the observed large-scale structure of the Universe. In contrast to hot dark matter, such as neutrinos, which remain relativistic until around matter-radiation equality and erase small-scale perturbations through free streaming, cold dark matter particles decouple from the primordial plasma and become non-relativistic by the epoch of matter-radiation equality, allowing small-scale density perturbations to survive and grow gravitationally. Early theoretical work and simulations showed that CDM model lead to hierarchical, bottom-up structure formation, where small objects form first and merge to build larger structures, which is consistent with the observed cosmic structures in our Universe. Large galaxy redshift surveys, such as the Sloan Digital Sky Survey (SDSS\cite{SDSS}) and the recent Dark Energy Spectroscopic Instrument (DESI\cite{DESI}) survey, revealed a large-scale web-like distribution of matter which is consistent with CDM predictions.

The original CDM paradigm was extended to the $\Lambda$CDM model following the discovery of cosmic acceleration from Type Ia supernovae observations in the late 1990s. Riess \cite{Riess1998} and Perlmutter \cite{Perlmutter}, based on observations of Type Ia supernovae as standard candles, showed that distant supernovae appeared dimmer than expected in a matter-dominated universe. This indicated that the expansion rate of the universe is accelerating which requires the presence of a dominant dark energy component-- parameterized as a cosmological constant $\Lambda$. The detailed historical steps leading to our understanding of dark matter and dark energy are shown in Table~\ref{table:CDMhistory}. 

\begin{table}[ht]
\centering
\caption{Key milestones in the Discovery of Dark Matter and Dark Energy}
\label{table:CDMhistory}
\begin{tabular}{cp{3cm}p{8cm}}
\hline
\textbf{Year} & \textbf{Reference} & \textbf{Discovery / Contribution} \\
\hline
1933 & Zwicky \cite{Zwicky} & First evidence for unseen matter: high galaxy velocity dispersion in the Coma cluster implies much more mass than luminous matter alone. \\
\hline
1973--1976 & Lea et al.~\cite{Lea}, Cavaliere \& Fusco-Femiano \cite{Cavaliere} & Detection of hot intracluster X-ray gas, requiring deeper gravitational potentials than provided by visible matter. \\
\hline
1978--1980 & Rubin et al.~\cite{Rubin} & Compelling evidence from flat galaxy rotation curves, pointing to extended dark matter halos. \\
\hline
1980s & Lynds \& Petrosian \cite{Lynds}, Soucail et al.~\cite{Soucail} & Discovery of giant arcs in clusters, early strong lensing evidence for dark matter. \\
\hline
1992 & Smoot et al.~\cite{COBE} & COBE detection of CMB anisotropies, confirming primordial density fluctuations and support for structure formation models based on dark matter. \\
\hline
1998--1999 & Riess et al.~\cite{Riess1998}, Perlmutter et al.~\cite{Perlmutter} & Discovery of cosmic acceleration from Type Ia supernovae, providing direct evidence for a nonzero cosmological constant ($\Lambda$). \\
\hline
2000 & Miller et al.~\cite{TOCO}, de Bernardis et al.~\cite{Bernardis}, Hanany et al.~\cite{MAXIMA} & TOCO, BOOMERanG and MAXIMA measurements of the first acoustic peak in the CMB angular power spectrum, providing strong evidence for a spatially flat Universe and greatly strengthening the $\Lambda$CDM paradigm. \\
\hline
2000s--2010s & WMAP \cite{WMAP}, Planck \cite{Planck} & High-precision measurements of the CMB power spectrum, tightly constraining cosmological parameters within the $\Lambda$CDM framework. \\
\hline
\end{tabular}
\end{table}

Building on these astrophysical evidences, the $\Lambda$ Cold Dark Matter ($\Lambda$CDM \cite{Peebles, Blumenthal, Davis}) model has established as the standard paradigm in modern cosmology, offering a robust and predictive framework for describing the formation and evolution of cosmic structures. In this model, the universe is primarily composed of dark energy ($\Lambda$) which dominates $\sim70\%$ of the total energy density of the Universe, and cold dark matter (CDM) $\sim27\%$, with baryonic matter constituting only a small fraction of the total energy density. The $\Lambda$CDM framework successfully explains a wide range of astrophysical and cosmological observations, from the large-scale distribution of galaxies to the anisotropies of the cosmic microwave background (CMB). CMB is the relic radiation from the hot, dense state of the early universe, emitted approximately 380,000 years after the Big Bang during the epoch of recombination. It encodes rich information about the initial density fluctuations that later grew into the large-scale structure we observe today. High-precision measurements of the CMB temperature and E-mode polarization power spectra (TT, TE, and EE) from the WMAP\cite{WMAP} and {\it Planck}\cite{Planck} missions have shown remarkable agreement with the predictions of the $\Lambda$CDM model. These constraints have allowed for precise determinations of cosmological parameters such as the Hubble constant ($H_0$), the matter density ($\Omega_m$), and the amplitude of primordial fluctuations ($\sigma_8$).

Despite its success, the $\Lambda$CDM model faces some challenges and tensions that motivate ongoing theoretical and observational efforts. One issue is the Hubble tension, which refers to the discrepancy between the locally measured value of the Hubble constant ($H_0$) using distance ladder techniques \cite{Riess} and the value inferred from CMB observations \cite{Planck}. In addition, several small-scale structure anomalies—such as the missing satellites problem, the core-cusp problem, and the too-big-to-fail problem—also show significant challenges to the standard $\Lambda$CDM model. These tensions may point to strong baryonic feedback processes, driven by supernovae or active galactic nuclei, which could reshape the internal structure of halos. Alternatively, they have motivated the exploration of alternative dark matter models—such as self-interacting dark matter (SIDM) which allows dark matter particles undergo self-scattering interactions. This self-interaction can transform CDM cuspy halo profiles into cored ones and reduce central densities, thereby offering a promising solution to the small-scale tensions between simulations and observations. 

Several comprehensive reviews of SIDM already exist, including those by Tulin \& Yu \cite{TulinYu} and Adhikari et al.~\cite{Adhikari2022}. This review focuses on the impact of dark matter self-interactions on the formation and evolution of cosmic structures, with particular emphasis on galaxy clusters, which provide some of the most powerful astrophysical laboratories for testing SIDM. We examine the theoretical motivation for SIDM and its current observational constraints. We begin by reviewing the general properties of cosmic structures within CDM in Section 2. Section 3 presents the key small-scale challenges faced by the CDM model. In Section 4, we discuss the motivation for SIDM as a promising alternative dark matter model, along with its theoretical predictions for cosmic structure formation. Section 5 summarizes the current observational constraints on the dark-matter self-interaction cross section. Finally, Section 6 concludes with a summary and  future aspects in the study of dark matter.

\section{Cosmic Structure with CDM }
\label{sc:CDM sims}

Peebles \cite{Peebles} first calculated a model in which the universe is dominated by weakly interacting, non-relativistic dark matter particles, seeded with adiabatic density perturbations and a scale-invariant primeval power spectrum of the form $P(k)\propto k$. Due to the suppression of perturbation growth on sub-horizon scales during the radiation-dominated era, and the subsequent growth of all linear modes in the matter-dominated regime, the initially scale-invariant power spectrum evolves: on small scales, or large
$k$, the matter power spectrum is suppressed, leading to $P(k)\propto k^{3}$, while on large scales it retains the original scale-invariant form
$P(k)\propto k$. The detailed calculation of the linear evolution of the CDM power spectrum was then provided by Blumenthal et al.\cite{Blumenthal1983} and Bond $\&$ Efstathiou \cite{BondEfstathiou}.
Peebles' model successfully explained the inhomogeneous mass distribution observed in the local Universe and predicted that primordial density fluctuations would leave observable imprints in the temperature anisotropies of the CMB. These predictions were later confirmed by high-precision CMB experiments, including COBE\cite{COBE}, TOCO\cite{TOCO}, BOOMERanG\cite{Bernardis},  {MAXIMA}\cite{MAXIMA}, WMAP\cite{WMAP}, and {\it Planck}\cite{Planck}.

In addition to theoretical developments, Peebles\cite{Peebles1973} also established the statistical framework for analyzing the large-scale distribution of galaxies via the two-point correlation function $\xi(r)$, describing the probability over random of finding two galaxies separated by a distance $r$ which then became a standard tool for testing cosmology.  Peebles shows that galaxy clustering follows a power-law form $\xi(r) \propto r^{-1.8}$. This provided one of the earliest empirical characterizations of cosmic structure, later interpreted in the context of CDM-based hierarchical clustering.

Soon after, Blumenthal et al. \cite{Blumenthal} developed a detailed theoretical model that incorporated the effects of dark matter on structure formation. Under the CDM hierarchical structure formation,  initial dark matter fluctuations collapse under gravity, grow nonlinearly, and virialize, leading to the formation of galaxies and clusters. Their model successfully explained several key observational properties, including the characteristic mass scale of galaxies, as well as empirical scaling relations such as the Faber–Jackson and Tully–Fisher relations. By adopting a cosmological model in which cold dark matter outweighs baryonic matter, they demonstrated that CDM provides a remarkably good match to the observed large-scale structure of the Universe. 

Building on the theoretical insights of CDM, Davis et al. \cite{Davis} conducted the first numerical simulations to investigate the nonlinear gravitational clustering of CDM particles. By adopting a simplified galaxy bias model—assuming that galaxies form preferentially in the high-density peaks of the initial fluctuation field—their simulations could reproduce observational statistics, including the galaxy two-point correlation function. In the same year, Frenk \cite{Frenk1985} demonstrated that CDM simulations naturally predict the presence of extended dark matter halos around galaxies, yielding flat rotation curves in the outer regions, in agreement with observations. These results provided strong support for the CDM paradigm.

\begin{figure*}[htbp]
 \centering
 \includegraphics[width=0.8\textwidth,trim={0.5cm 0cm 0cm 0cm},clip]{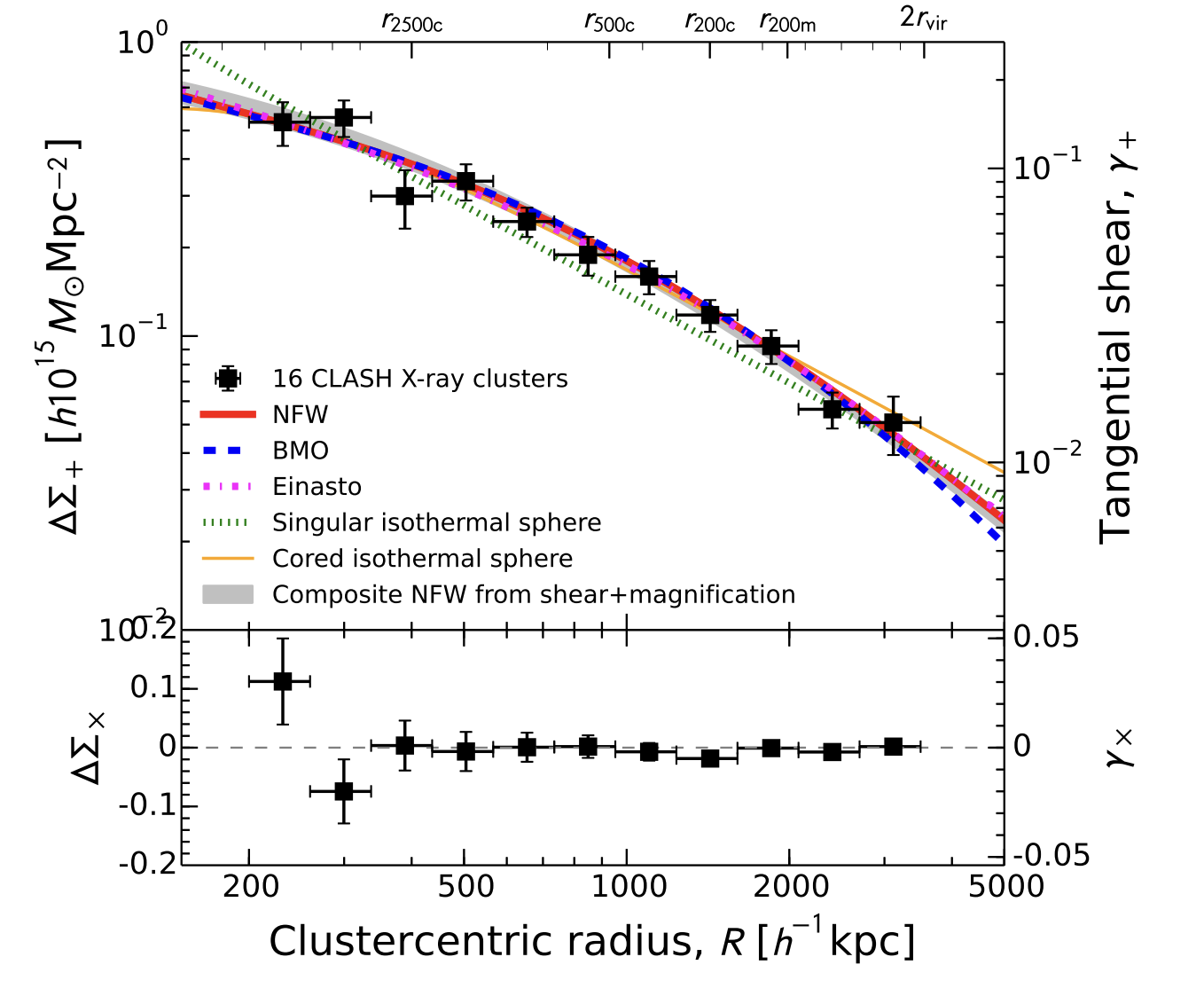}
\caption{Upper panel: The average tangential shear profile obtained by stacking a subsample of 16 X-ray-selected clusters from the CLASH survey. Colored lines represent different best-fit halo models. Their result shows NFW provides good fit to the observed data. 
Lower panel: The $45^\circ$-rotated ($\times$) shear component, consistent with a null signal. Reprinted from Umetsu et al. \cite{Umetsu2014}.
 }
 \label{fig:Umetsu2014}
\end{figure*}

By the early 1990s, further advancements in computational techniques enabled more detailed simulations of dark matter halo formation. Large cosmological simulations demonstrated that dark matter collapses into halos with well-defined density profiles. The spherically averaged mass distribution was found to follow a universal form, known as the Navarro–Frenk–White (NFW) profile,\cite{Navarro1996} characterized by a cuspy inner density distribution:
\begin{equation}
     \rho(r) = \frac{\rho_s}{\frac{r}{r_s}(1 + \frac{r}{r_s})^2}, 
\end{equation}
where $\rho_s$ is the characteristic density and $r_s$ is a characteristic scale radius at which the logarithmic density slope, $\alpha=d\ln{\rho}/d\ln{r}$, equals -2. The NFW profile shows that dark matter halos exhibit a cuspy inner density structure, with the density scaling as $\rho(r) \propto r^{-1}$ near the halo center and transitioning to $\rho(r) \propto r^{-3}$ at large radii. 
%Subsequent studies, such as those by \cite{Moore1998, Fukushige}, proposed an even steeper inner slope in the central regions of halos, with $\alpha\sim-1.5$.
Subsequent high-resolution simulations \cite{Ishiyama2014} suggested that the earliest, low-mass halos in the universe might exhibit even steeper inner slopes (e.g., $\alpha \sim -1.5$), while larger halos forming through mergers and accretion relax toward shallower profiles. This universality of halo profiles has been confirmed across a wide range of masses, from dwarf galaxies to galaxy clusters, and is also supported by gravitational lensing \cite{Umetsu2014} that probe the mass distribution in halos, as shown in Figure~\ref{fig:Umetsu2014}. 

Some high-resolution simulations \cite{Merritt, Dutton2014} have shown that the NFW profile, while broadly successful, may not capture the detailed shape of the inner halo density.  To address these deviations, alternative parametric models have been proposed that offer greater flexibility in fitting halo profiles. These include the generalized NFW (gNFW) profile \cite{Zhao1996} and the Einasto profile \cite{Einasto}, both of which introduce additional degrees of freedom to better describe the spherically averaged density structure. For example, Einasto profile is defined as 

\begin{equation}
\rho_{\mathrm{Einasto}} = \rho_s \exp \left\{ -\frac{2}{\alpha_E} \left[ \left( \frac{r}{r_s} \right)^{\alpha_E} - 1 \right] \right\}, \tag{B4}
\end{equation}
where $\alpha_E$ is the shape parameter describing the steepness of the logarithmic slope. An Einasto profile with $\alpha_E \sim 0.18$ closely resembles the NFW profile at a given concentration.

With the advancement of computational resources, high-resolution and large-scale simulations have been developed to study structure formation within the $\Lambda$CDM framework. For instance, the Millennium Simulation employs $10^{10}$ particles to trace the evolution of dark matter in a cubic volume of $500\,h^{-1}\,\mathrm{Mpc}$ per side, achieving a spatial resolution of $5\,h^{-1}\,\mathrm{kpc}$. In addition to tracking dark matter dynamics, semi-analytic models of galaxy formation are implemented to account for baryonic processes such as gas cooling, star formation, and the growth of supermassive black holes.

Although CDM N-body simulations have provided crucial insights into the formation and evolution of dark matter halos, recent state-of-the-art simulations have significantly improved both particle and spatial resolution, and more importantly, the treatment of baryonic physics, including galaxy formation and feedback mechanisms. 
These baryonic processes from supernovae and active galactic nuclei, play crucial roles in shaping the observable Universe. Semi-analytic models \cite{White1991} and cosmological hydrodynamical simulations such as EAGLE \cite{Schaye2015}, IllustrisTNG \cite{Nelson2018}, and MillenniumTNG \cite{Pakmor2023} employ advanced hydrodynamical models to simulate the co-evolution of dark matter and baryons. These simulations shows baryonic processes can alter the distribution of dark matter halos, for instance by flattening inner density profiles through energetic feedback, thereby potentially resolving tensions between CDM predictions and certain small-scale observations which we will discuss more in Section~\ref{sc:small-scale}. 

It is worth noting that while the CDM paradigm has been remarkably successful in explaining large-scale cosmic structure, alternative theories have been proposed to address discrepancies at galactic scales, during the same period. One such alternative is Modified Newtonian Dynamics (MOND), proposed by Milgrom \cite{Milgrom},  suggesting that the observed flat rotation curves of galaxies could be explained by modifying Newtonian gravity at low accelerations rather than introducing an unknown matter component. MOND modifies Newton’s second law as
\begin{equation}
   \mu\left(\frac{a}{a_0}\right) a = -\nabla\Phi_{N},
\end{equation}
where $a_0\approx1.2\times10^{-10}\rm m/s^2$, and $\mu(x)$ is an interpolating function satisfying 
$\mu(x)\approx x$ in the deep-MOND regime ($x\ll1$), and $\mu(x)\approx 1$ in the Newtonian regime ($x\gg1$).
Although MOND has proven successful in fitting galactic rotation curves, it faces significant challenges in explaining large-scale cosmic structures, such as galaxy clusters, and the CMB anisotropies. To address these limitations, some extended formulations of MOND have been proposed, including extended MOND \cite{Zhao2012}, TeVeS \cite{Bekenstein2005} and relativistic MOND \cite{Skordis2021}. Nevertheless, the cold dark matter (CDM) paradigm remains the most widely accepted and empirically successful framework for explaining the broad range of astrophysical and cosmological observations.

For a more detailed overview of structure formation in the CDM context, we refer the reader to Chapter 14 of GR100, authored by Marc Davis.

\section{Small-Scale Structure Challenges }
\label{sc:small-scale}
While CDM simulations have been successful in describing the large-scale structure of the Universe, several observational discrepancies exist on small scales that may pose challenges to the standard $\Lambda$CDM
model. These tensions may suggest that either additional physical processes need to be incorporated into current models or an alternative dark-matter model may be required to explain small-scale phenomena.

\subsection{The Core-Cusp Problem}
As discussed earlier, the cuspy NFW profile predicted by the $\Lambda$CDM model effectively characterizes the density distribution of dark matter halos in cosmological simulations. However, observational studies \cite{Moore1994, Moore1999} of dwarf galaxies and low-surface-brightness galaxies, which are typically dominated by dark matter, have showed linearly rising rotation curves in their inner regions. These kinematic features imply central density profiles that are substantially shallower than the steep, cuspy inner slopes predicted by the NFW model. Instead of exhibiting the characteristic $\rho \propto r^{-1}$ cusp, these galaxies tend to show flat, constant-density cores in their central regions. 
This discrepancy, commonly referred to as the core-cusp problem, has been widely discussed in the literature \cite{Bullock,Popolo} and remains one of the challenges of $\Lambda\rm CDM$.  
\iffalse
To better fit the observed cored profiles, alternative empirical model have been proposed, e.g. Burkert \cite{Burkert1995} profile defined as 
\begin{equation}
     \rho(r) = \frac{\rho_0}{(1 + r/r_c)\left[1 + (r/r_c)^2\right]},
\end{equation}
where $\rho_0$ and $r_c$ are the central density and core radius, 
introduces a central constant-density core and provides better agreement with the observed rotation curves of dwarf galaxies. 
\fi

For example, with higher-resolution rotation curve data, studies \cite{Oh2011, Oh2015} from the HI Nearby Galaxy Survey (THINGS) and LITTLE THINGS have shown that the logarithmic inner slopes of dwarf galaxies are approximately $\alpha \sim -0.3$. This value significantly deviates from the cuspy dark matter distribution with $\alpha \sim -1$ predicted by dark matter-only $\Lambda$CDM simulations. Instead, the observed shallower slopes are more consistent with predictions from hydrodynamic simulations incorporating baryonic processes, such as gas cooling, star formation, and feedback mechanisms.

One proposed solution to this discrepancy between CDM predictions and observations  is the formation of dark matter cores through baryonic processes. Non-gravitational processes—such as energy injection from supernovae and active galactic nuclei (AGN)—can induce fluctuations in the gravitational potential, redistributing mass and thereby modifying the inner structure of dark matter halos. Using high-resolution hydrodynamical simulations, Governato et al. \cite{Governato2010} demonstrated that strong supernova-driven outflows can effectively remove low-angular-momentum gas, suppressing bulge formation and reducing the central dark matter density. This mechanism naturally produces cored dark matter profiles with typical sizes of a few kiloparsecs. More recent state-of-the-art cosmological hydrodynamical simulations, including GASOLINE \cite{Governato2010, Tollet2016}, FIRE \cite{Chen2015, Hopkins2018}, EAGLE \cite{Llambay2019}, and IllustrisTNGhave further explored core formation under various feedback models. While many of these simulations successfully reproduce cored profiles in low-mass galaxies, they often produce different inner dark matter density profiles due to subgrid physics modelling. For comprehensive studies and reviews on how baryon physics lead to core formation, we refer readers to literature \cite{Blok2010,  Tollet2016, Hopkins2018, Popolo2021}. An alternative solution involves introducing new dark matter models that alter the internal structure of halos. These will be discussed in detail in Section~4.

\subsection{The Missing Satellites Problem}
In the $\Lambda$CDM model, cosmic structures grow hierarchically through the mergers and accretions of smaller dark matter halos. This framework predicts a vast amount of substructures, or subhalos, embedded within larger host halos, including those around Milky Way-like galaxies \cite{Kauffmann1993, Moore1999ApJ}. High-resolution $N$-body simulations \cite{Springel2008} predict hundreds to thousands of dark matter subhalos within Milky Way-like galaxies.

However, observations of the Local Group show a significant discrepancy: the number of detected dwarf satellite galaxies is roughly an order of magnitude lower than the number of subhalos predicted by $\Lambda$CDM simulations \cite{Klypin1999}. This discrepancy, commonly referred to as the “missing satellites problem”, raises important questions about the nature of dark matter and whether baryonic processes, such as feedback, suppress star formation in low-mass halos \cite{Brooks2013}, making many subhalos dark and thus difficult to detect observationally.

Recent deep-imaging surveys have aimed to uncover these hidden satellite populations. For example, Homma et al.\cite{Homma2024} used data from the Hyper Suprime-Cam (HSC) Subaru Strategic Program (SSP) to discover new candidate satellite galaxies of the Milky Way. Accurately characterizing the number of satellite galaxies remains challenging due to observational limitations. Next-generation facilities, such as the Vera C. Rubin Observatory will provide deeper, high-resolution, and wide-field imaging for bridging the gap between observed and predicted subhalo abundances.

\subsection{The Too-Big-to-Fail Problem}
Boylan-Kolchin et al. \cite{TBTF} introduced the Too-Big-to-Fail (TBTF) problem, highlighting another discrepancy between $\Lambda$CDM simulations and observations of the Milky Way's satellite galaxies. Using high-resolution DM-only simulations of galactic haloes of the Aquarius project \cite{Springel2008}, they found that the most massive subhalos, which are expected to host the brightest satellites, have central densities and maximum circular velocities ($V_{\rm max} > 30~\rm km/s$) that are significantly higher than those inferred from kinematic observations. However, the measured maximum circular velocities of the Milky Way’s dwarf spheroidal galaxies suggest that only about three satellites exhibit $V_{\rm max}$ values above 30 km/s. This infer that most of Milky Way's satellites reside in halos with much lower central densities. This mismatch implies that the most massive subhalos predicted by $\Lambda$CDM simulations are too dense to host any of the known satellites, leading to the ``too big to fail" problem.

Several solutions have been proposed to address this discrepancy.  One possibility is to adopt a lower total mass for the Milky Way halo \cite{Wang2012, Cautun2014}. A reduced halo mass would naturally yield fewer massive subhalos, thereby lowering the predicted $V_{\rm max}$ of satellites. The second approach involves baryonic feedback processes, where supernova-driven outflows and tidal interactions can redistribute mass within subhalos, flattening their central density profiles and preventing the formation of overly dense cores \cite{Brooks}.

Alternatively, modifications to the nature of dark matter have been explored. Self-interacting dark matter (SIDM) models, for instance, introduce particle self-scatterings that can redistribute energy within halos, leading to the formation of isothermal cores and reducing central densities \cite{Zavala2013, Cline2014, Vogelsberger2016}. Warm dark matter (WDM) models suppress small-scale structure due to free-streaming effects in the early universe, resulting in a reduced abundance of small subhalos and delayed collapse, which in turn lowers their concentration \cite{Lovell2012, Lovell2017}.

%These alternative models provide viable mechanisms to alleviate the TBTF problem while maintaining consistency with large-scale structure formation. A more detailed comparison of these models is discussed in Section~\ref{sec:SIDM}.
%https://arxiv.org/pdf/1602.05957

\iffalse
\subsection{Impact from Baryons}
Feedback from supernovae and AGN can flatten density cusps, reducing tensions between CDM and observed rotation curves.

Simulations predict a vast number of subhalos around Milky Way-like galaxies, but many may be destroyed by tidal effects or baryonic feedback.

serveal different solution --> focus on SIDM 
\fi

\section{Self-Interacting Dark Matter-- SIDM}
Although various approaches have been proposed to address the small-scale challenges of the Cold Dark Matter (CDM) paradigm, this review focuses on Self-Interacting Dark Matter (SIDM) as a compelling alternative.

SIDM was first proposed by Spergel and Steinhardt \cite{Spergel2000}, introducing the idea that dark matter particles can remain cold, but undergo elastic self-scattering with a non-negligible cross-section, typically characterized by the scattering cross-section per unit mass $\sigma/m$. They suggested that a cross-section in the range \( \sigma/m \sim 0.1\text{--}100\, \mathrm{cm}^2/\mathrm{g} \) %\footnote{For reference, the  electron–electron (Møller) scattering cross section is $d\sigma/d\Omega \propto 1/(E^2)\propto 1/(v^4)$. At $E =$ GeV scale, $\sigma$ has an order of $10^{-8}$ barn, where $1\,\mathrm{barn} \;\approx\; 10^{-24}\,\mathrm{cm^2}$.} 
could lead to observable effects on the internal structure of dark matter halos. These self-interactions allows energy and momentum to be redistributed in the inner regions of dark matter halos. As a result, the central density profiles become shallower, often forming isothermal cores, in contrast to the cuspy profiles predicted by collisionless CDM. SIDM largely preserves the large-scale predictions of the CDM paradigm, including the halo and subhalo abundance, while altering the internal density and kinematic structure of dark matter halos through particle self-scattering. This makes SIDM a simple and appealing extension of CDM, as it can explain both the large-scale structure of the Universe and the small-scale features inside halos. This core formation helps to alleviate several small-scale discrepancies between CDM predictions and astrophysical observations, such as the core-cusp and too-big-to-fail problems. 

In SIDM model, the local scattering rate is given by  
\begin{equation}
\label{eq:scatterrate}
\Gamma = \rho_{\rm dm}(r)\, v(r)\, \frac{\sigma}{m} \approx 0.1\, \mathrm{Gyr}^{-1} \times \left( \frac{\rho_{\rm dm}}{0.1\, M_{\odot}/\mathrm{pc}^3} \right) \times \left( \frac{v}{50\, \mathrm{km/s}} \right) \times \left( \frac{\sigma/m}{1\, \mathrm{cm}^2/\mathrm{g}} \right),
\end{equation}
where \( m \) is the dark matter particle mass, \( v \) is the relative velocity, and \( \rho_{\rm dm} \) is the local dark matter density. For sufficiently large values of \( \sigma/m \), the scattering probability becomes significant over the lifetime of a halo. When \( \Gamma\, t_{\rm age} \sim 1 \), self-interactions occur frequently enough to noticeably alter the halo's internal structure. As shown in Equation~\eqref{eq:scatterrate}, the scattering rate also depends on the local dark matter density. Therefore, at sufficiently large radii where the density is low, SIDM halos resemble those in CDM. On the other hand, massive galaxy clusters—with high \( \rho_{\rm dm} \) and the typical order of velocity \( v\sim1000\,\rm km/s \)—provide ideal environments to search for signatures of dark matter self-interactions. Observations of galaxy clusters generally place upper limits in the range
$\sigma/m \lesssim 0.1$--$1\,\mathrm{cm}^2\,\mathrm{g}^{-1}$,
depending on the observational probe and the assumptions adopted in the analysis. We discuss these constraints in detail in Section~5. However, in order to address small-scale discrepancies in galaxies, a larger value of \( \sigma/m \gtrsim 1\, \mathrm{cm}^2/\mathrm{g} \) is often required \cite{Dave2001,Kaplinghat2016}.

This tension between constraints from galaxy and cluster scales motivates an extension to velocity-dependent SIDM models, where the cross-section \( \sigma(v)/m \) decreases with increasing relative velocity. This allows SIDM models to produce large self-interactions in low-velocity systems (e.g., dwarf galaxies) while suppressing interactions in high-velocity systems (e.g., galaxy clusters), thus satisfying constraints across a wide range of mass scales.

From the particle-physics perspective, SIDM can be realized in a wide variety of models, many of which invoke a hidden sector containing new mediator particles, such as dark photons or light scalars. In these models, dark matter interacts via a force carrier that does not couple directly to Standard Model particles. The resulting self-scattering is usually described by Yukawa potential, 
\begin{equation}
V(r) = \pm \frac{\alpha_{\chi}}{r} e^{-m_\phi r},
\end{equation}
where $\alpha_{\chi}$ is the coupling strength and $m_\phi$ is the mediator mass. By using Born approximation \cite{Jelley1990}, the differential cross section can be expressed by \cite{IbeYu}
\begin{equation}
\frac{d\sigma}{d\Omega} = \frac{\alpha_\chi^2}{\left( m_\phi^2+ m_\chi^2 v^2 \sin^2\left( \frac{\theta}{2} \right) \right)^2},
\end{equation}

With a massive mediator $m_\phi\gg m_\chi v$, dark matter scatterings result in a short-range force and typically isotropic scattering. In contrast, a light mediator $m_\phi\ll m_\chi v$ gives rise to a long-range force, where the self-interaction cross-section depends on the relative collision velocity and scattering angle, which leads to velocity-dependent and anisotropic scattering.  

The self-interaction cross-section inferred from astrophysical observations can provide valuable constraints on the fundamental properties of dark matter, including its mass, the mass of the mediator, and the coupling strength between them \cite{Kaplinghat2016}. For example, Kaplinghat et al. (2016) \cite{Kaplinghat2016} estimated the  dark matter mass of $\sim$ 15 GeV and the mass of a dark photon $\sim$ 17 MeV  by considering the core sizes observed in galaxy clusters \cite{Newman2013}, low surface brightness galaxies \cite{Kuzio de Naray 2008}, and dwarf galaxies \cite{Oh2011}.

To understand how dark matter self-interactions affect the structure and evolution of halos, both semi-analytic models and numerical simulations have been developed. Semi-analytic approaches provide a physically intuitive description of the thermal evolution of SIDM halos and help identify the key processes governing core formation and gravothermal collapse. We first review these theoretical frameworks before discussing the results of numerical simulations.

\subsection{Semi-analytic Models}
Semi-analytic models play a crucial role in advancing our understanding of SIDM. These models provide complementary insights by capturing the essential physics of dark matter self-interactions in a computationally efficient way. Balberg, Shapiro, and Inagaki \cite{Balberg2002} developed the first semi-analytic study of the dynamical evolution of isolated SIDM halos. They examined the dynamical evolution of isolated SIDM halos by assuming spherical symmetry and self-similar evolution, and employed a gravothermal fluid formalism to model heat transport driven by elastic dark matter self-interactions. Their formulation relies on a set of fundamental equations that describe mass conservation, hydrostatic equilibrium, and the first law of thermodynamics, as follows

\begin{align}
\frac{\partial M}{\partial r} &= 4\pi r^2 \rho, \\
\frac{\partial}{\partial r} (\rho v^2) &= -\rho \frac{GM}{r^2}, \\
-\frac{1}{4\pi r^2} \frac{\partial L}{\partial r} &= \rho v^2 \frac{\partial}{\partial t} \ln\left( \frac{v^3}{\rho} \right),
\end{align}
where the time derivatives denotes a Lagrangian derivative.
To describe thermal conduction, Balberg et al. \cite{Balberg2002} developed a flux equation that smoothly interpolates between the short mean free path (smfp) and long mean free path (lmfp) regimes

\begin{equation}
    \frac{L}{4\pi r^2} = -\frac{3}{2} ab v \sigma \left[ a \sigma^2 + \frac{4\pi G}{\rho v^2} \right]^{-1} \frac{\partial v^2}{\partial r},
\end{equation}
where the first term dominates in the smfp regime—when interactions are frequent and short-ranged, while second term corresponds to the lmfp limit when particles travel farther between collisions. These two regimes represent different modes of heat conduction and mass transfer between the halo core and its extended envelope. This framework describes the evolution of a spherical, isolated SIDM halo. Initially, thermal conduction drives energy inward, leading to the formation of a central isothermal core. Over time, as heat flows outward, the core gradually contracts, ultimately resulting in gravothermal core collapse. Although this model shows good agreement with N-body simulations of isolated halos, Koda et al. \cite{Koda2011} reported discrepancies of about 20$\%$ in both the central density and the collapse rate when compared with N-body simulation results. Therefore, although the conducting fluid model successfully captures the qualitative evolution of SIDM halos, it does not fully reproduce the detailed structure and evolution observed in cosmological N-body simulations.

Alternatively, Kaplinghat et al.~\cite{Kaplinghat2014, Kaplinghat2016} developed a semi-analytic framework to model SIDM halos based on the isothermal Jeans model. Their approach is based on the fact that in the inner regions of SIDM halos, where the scattering rate is highest, frequent interactions thermalize the dark matter particles and produce an isothermal core. In this regime, the density profile satisfies the Jeans–Poisson equations under the assumptions of isotropy and spherical symmetry:

\begin{align}
\frac{d}{dr} (\rho_{\rm dm} \sigma_0^2)  &= -\rho_{\rm dm} \frac{d\Phi}{dr}, \\
\frac{1}{r^2} \frac{d}{dr} \left( r^2 \frac{d\Phi}{dr} \right) &= 4\pi G (\rho_{\rm dm} + \rho_{\mathrm{b}}),
\end{align}
where $\Phi$ is the total gravitational potential from both dark matter and baryons, $\rho_{\rm dm}$ is the dark matter density, $\rho_{\rm b}$ 
is the baryonic matter density and $\sigma_0$ is the one-dimensional velocity dispersion, assumed to be constant within the isothermal core.
Combining the above equations yields \cite{Kaplinghat2014}:
\begin{equation}
\label{eq:iso}
\frac{\sigma_0^2}{r^2} \frac{d}{dr} \left( r^2 \frac{d \ln \rho_{\mathrm{dm}}(r)}{dr} \right) 
= -4\pi G \left[ \rho_{\mathrm{dm}}(r) + \rho_{\mathrm{b}}(r) \right],
\end{equation}
which governs the dark matter density profile in the isothermal core. The solution to Equation~\ref{eq:iso} provides the so-called isothermal profile, 
$\rho_{\rm iso}(r)$, that characterizes the inner region of an SIDM halo. At larger radii, where the local dark matter density is lower and the scattering rate drops, self-interactions become rare and the system transitions to a collisionless regime. In this outer region, the density profile reverts to that of cold dark matter, typically well-described by the NFW profile. The full dark matter density profile, $\rho_{\rm dm}(r)$, can be modeled as a combination of a collisional isothermal core and a collisionless outer halo following the NFW profile:
\begin{equation}
\rho_{\mathrm{dm}}(r) = 
\begin{cases}
\rho_{\mathrm{iso}}(r), & r < r_1 \\
\rho_{\mathrm{NFW}}(r), & r > r_1
\end{cases}.
\end{equation}
Here, $r_1$ marks the transition radius at which the effects of self-interactions become negligible. This boundary is defined by the condition $\Gamma(r_1)\, t_{\rm age} \sim 1$, indicating that, on average, each dark matter particle has undergone approximately one collision over the cluster lifetime. Robertson et al.~\cite{Robertson2021} demonstrated that this isothermal Jeans model provides an accurate description of simulated SIDM density profiles, as seen in hydrodynamical simulations from both the BAHAMAS \cite{McCarthy2017, McCarthy2018} and EAGLE \cite{Schaye2015} projects.

Recently, Jiang et al.~\cite{Jiang2023} improved upon the isothermal Jeans model by incorporating a prescription for adiabatic halo contraction, enabling a more self-consistent treatment of baryonic effects. Their updated framework exhibits good agreement with cosmological SIDM simulations, accurately capturing the halo structure throughout the core-forming phase and up to the onset of gravothermal core collapse. By comparing this approach with the gravothermal fluid formalism \cite{Koda2011, Nishikawa2020}, they found that the isothermal Jeans model provides a closer match to simulation results during the early, core-forming stage. In contrast, the gravothermal fluid model is better suited for describing the later core-collapse phase, highlighting the complementary nature of the two approaches. Zhong et al.~\cite{Zhong2023} extended the analytical gravothermal fluid model to include the influence of a static baryonic potential. Their framework, calibrated with N-body simulations, demonstrates that baryonic components accelerate core evolution, shortening the timescales for both maximum core expansion and core collapse. The model also accurately tracks key halo properties, such as central dark matter density, core size, and velocity dispersion. As pointed out by Zhong et al.~\cite{Zhong2023}, a natural extension is to incorporate a time-varying baryonic potential, informed by hydrodynamical SIDM simulations. This would allow a more realistic modeling of core evolution by directly accounting for the evolving baryonic distribution.

%Based on these simulation results, recent studeis have developed some analytic or parametric model to model the evolution of SIDM halos and provide trainsition form CDM NFW profile to SIDM core density profile: https://arxiv.org/pdf/2305.16176

\subsection{Simulations}
While semi-analytic models provide valuable physical insight, numerical simulations are required to capture the full non-linear evolution of SIDM halos in a cosmological context and make detailed observational predictions \cite{Yoshida2000,Dave2001, Moore2000, Rocha2013, Peter2013, Elbert2015, Brinckmann2018}. Most of these simulations assume velocity-independent and isotropic scattering, characterized by a constant cross-section per unit mass, $\sigma/m$. For example, the first cosmological SIDM simulations by Dave et al. \cite{Dave2001} explored the impact of self-interactions on galaxy formation and found that cross-sections in the range $\sigma/m=0.1-10\,\rm cm^2/g$ could reproduce the observed core densities of dwarf galaxies. In contrast, Yoshida et al.\cite{Yoshida2000} focused on massive cluster-scale halos and demonstrated that a simulation with $\sigma/m=0.1\,\rm cm^2/g$ provides a good match to the core radius observed in the galaxy cluster CL 0024+1654. These results suggest that a single, velocity-independent value of \(\sigma/m\) cannot simultaneously explain observations across all halo mass scales. Consequently, more recent studies have motivated the use of velocity-dependent SIDM models, where the effective cross-section decreases with increasing relative velocity \cite{Vogelsberger2012}.

For details on the implementation of self-interactions in simulations, we refer readers to the original studies and the review by Tulin and Yu \cite{TulinYu}. Here we focus on the main physical effects gained from SIDM simulations and their implications for the structure of dark matter halos.

\subsubsection{Impact on Halo Structure}
\label{sec:density}
In SIDM halos, frequent elastic self-interactions in the dense inner regions cause dark matter particles to  exchange energy and momentum. This process drives the inner halo toward thermal equilibrium, producing two closely related structural changes: the formation of a lower-density, approximately isothermal core and a reduction in halo triaxiality. At larger radii, where the dark matter density and thus the scattering rate is lower, the density profile of SIDM halos gradually transitions back to the cuspy form described by the NFW profile.

\begin{figure*}[htbp]
 \centering
 \includegraphics[width=1\textwidth,trim={0.5cm 0cm 0cm 0cm},clip]{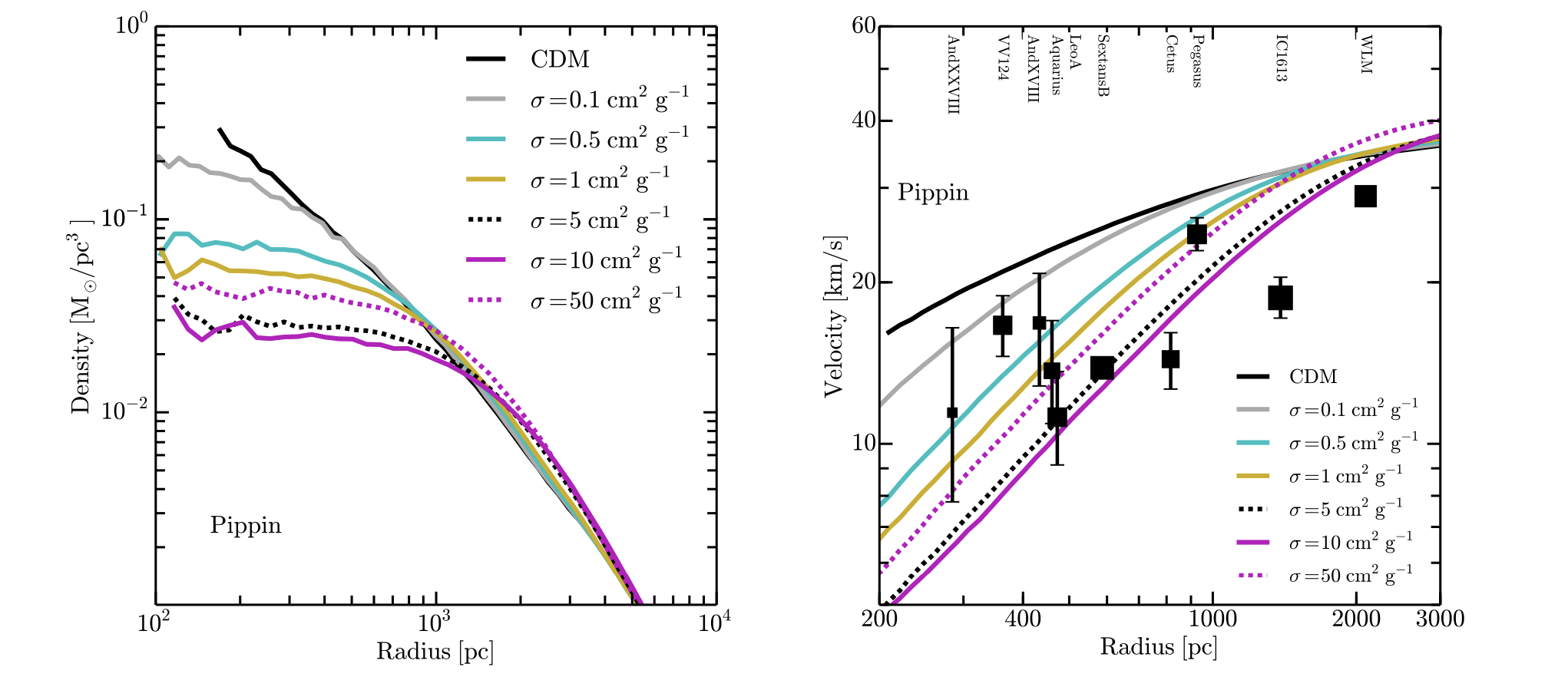}
\caption{Left: Dark matter density profiles for low-mass halos from SIDM N-body simulations with various values of  $\sigma/m$. Right: Corresponding circular velocity profiles. Black data points represent observed stellar circular velocities measured at the half-light radii of dwarf galaxies in the Local Group. Reprinted from Elbert et al. \cite{Elbert2015}.
 }
 \label{fig:Elbert2015}
\end{figure*}

For galaxy-scale halos, self-interaction cross sections of $\sigma/m \gtrsim 0.5\, \mathrm{cm}^2/\mathrm{g}$ have been shown to produce rotation curves consistent with observations of dwarf and low surface brightness galaxies \cite{Dave2001, Elbert2015}. Figure~\ref{fig:Elbert2015} illustrates the density profiles of halos with masses around $10^{10} M_{\odot}$ from SIDM DM-only simulations presented in Elbert et al.\cite{Elbert2015}. The results demonstrate that increasing $\sigma/m$ generally reduces the central dark matter density and leads to the formation of a constant-density core. However, for sufficiently large cross-sections ($\sigma/m\sim50\,\mathrm{cm}^2/\mathrm{g}$), the system enters a core-collapse regime and central density to rise again, although it remains shallower than the steep cusps predicted by collisionless CDM.

The same physical mechanism operates in cluster-scale halos. Yoshida et al. \cite{Yoshida2000} and Rocha et al. \cite{Rocha2013} performed cosmological simulations focusing on massive halos with $M \sim 10^{14} - 10^{15}M_{\odot}$. Similarly to galaxy-scale studies, with increasing $\sigma/m$, a larger and lower cental density core will form in the inner region (100-200 kpc) of the galaxy cluster. They showed that SIDM models with cross-sections $\sigma/m \sim 0.1 \,\mathrm{cm}^2/\mathrm{g}$ are consistent with the observed shallow density cores in galaxy clusters. Further observational constraints on the SIDM cross-section at cluster scales will be discussed in Section~5.

\begin{figure*}[htbp]
 \centering
 \includegraphics[width=1\textwidth,trim={0.5cm 0cm 0cm 0cm},clip]{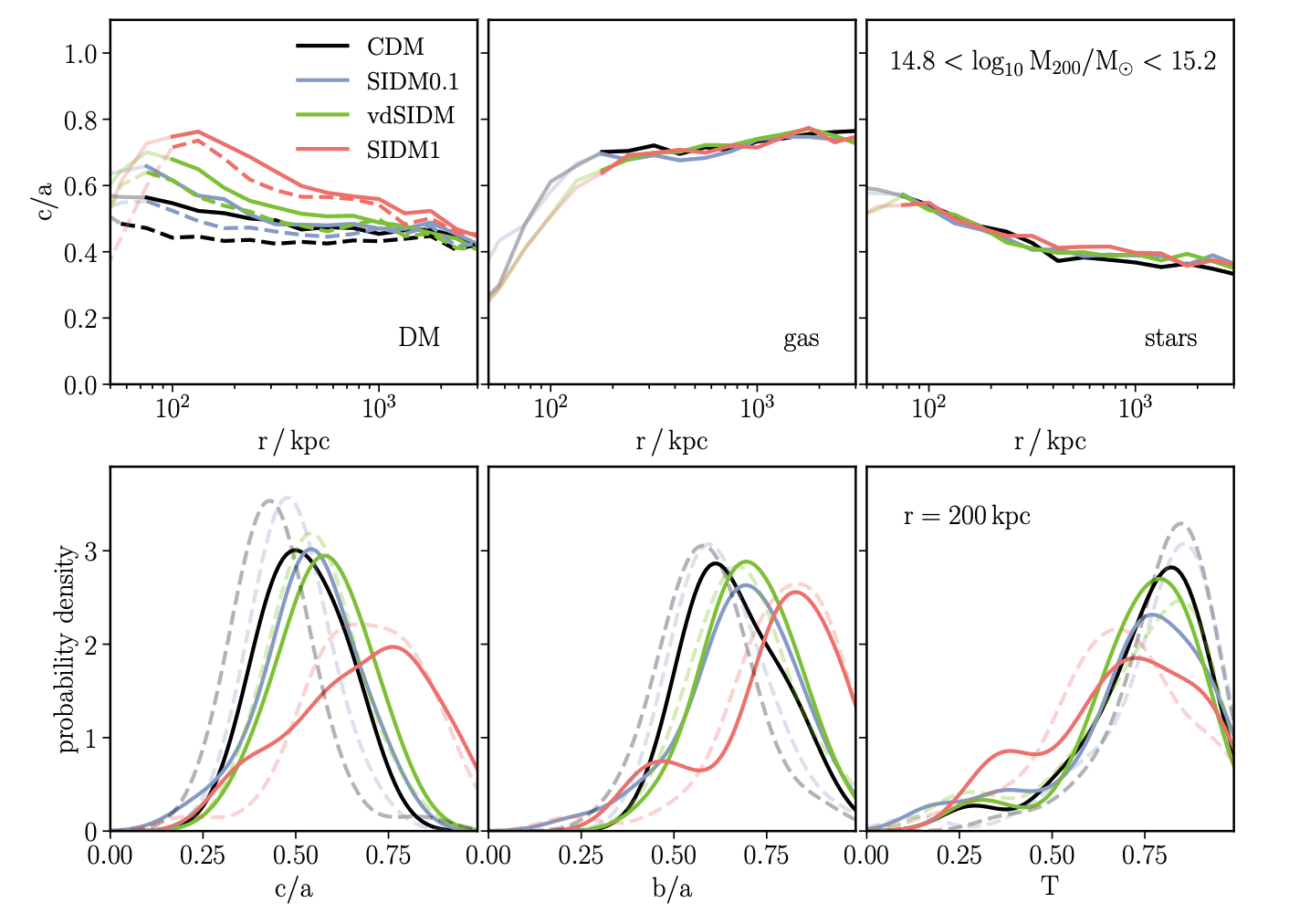}
\caption{
Top: Solid lines show the minor-to-major axis ratios for dark matter, gas, and stars from the BAHAMAS hydrodynamical simulations, while dashed lines represent the corresponding results from dark matter-only simulations. Bottom: The corresponding distributions of dark matter axis ratios at $r = 200$~kpc, where T is he triaxiality parameter defined as $T = (a^2 - b^2)/(a^2 - c^2)$. The distribution indicates a preference for prolate halos with $T \geq 2/3$. Reprinted from Robertson et al.~\cite{Robertson2019}.
}
 
 \label{fig:Robertson2019}
\end{figure*}

Self-interactions also modify halo morphology. \cite{Dave2001, Yoshida2000, Peter2013}.
In contrast to the triaxial shapes commonly found in collisionless CDM halos, SIDM tends to isotropize particle velocities through scattering, leading to more spherical halo shapes, especially in the inner regions with radius dependence. 
Peter et al.\cite{Peter2013} used cosmological simulations to study halo shapes across a broad mass range from $M \sim 10^{11}$ to $10^{14}M_{\odot}$. They showed that self-interactions substantially round halo cores. For a cross-section of $\sigma/m = 1\,\mathrm{cm}^2/\mathrm{g}$, the minor-to-major axis ratio in the inner regions can reach $c/a \sim 0.8$, while for $\sigma/m = 0.1\,\mathrm{cm}^2/\mathrm{g}$, the typical value is $c/a \sim 0.5$–$0.6$. These are  rounder than the typical axis ratios in CDM halos, which are usually $c/a \sim 0.45$–$0.5$. Similar trends have also been observed in Brinckmann et al.\cite{Brinckmann2018} and hydrodynamical cosmological simulations, such as those from the BAHAMAS-SIDM suite \cite{Robertson2019}, as shown in Figure ~\ref{fig:Robertson2019}.

These changes in density structure and halo morphology provide complementary observational signatures of SIDM. Density cores are most prominent in the central regions, while changes in halo shape can extend to larger radii and can be probed observationally through gravitational lensing analyses, which provide constraints on the projected ellipticity of dark matter distributions in galaxies and clusters. However, lensing measurements are subject to projection effects along the line of sight. Since lensing is sensitive to the integrated mass distribution, contributions from the more triaxial outer halo can add contribution to the rounder shapes in the inner regions--thereby weakening the constraining power on $\sigma/m$ \cite{Peter2013, Robertson2023}.

\subsubsection{Hydrodynamical Simulations}
Although dark matter-only simulations have established the basic structural effects of self-interactions, realistic predictions for galaxies and galaxy clusters require the inclusion of baryonic physics. Processes such as star formation, supernova feedback, and active galactic nucleus (AGN) feedback can significantly alter halo structure and therefore influence the observational signatures of SIDM. Therefore, implementing dark matter self-interactions into cosmological hydrodynamical simulations that self-consistently include baryonic physics is essential. 

One of the first such studies was carried out by Vogelsberger et al., who examined the structure of two dark matter-dominated dwarf galaxies with $M_*/M_{\rm DM} \leq 0.15$. They found that the effect of baryonic feedback on the dark matter density profile was minimal, likely due to the use of a relatively smooth star formation model. The resulting core sizes and central densities were similar to those found in DM-only SIDM simulations. They observed that dark matter self-interactions did have a significant impact on the baryonic component. Within the central kiloparsec, self-interactions reduced the stellar and gas densities, velocity dispersions, and gas temperatures by up to $\sim 50\%$. These results suggest that the stellar distribution within the core radius could serve as a potential observational probe of SIDM.

Fry et al. also conducted the first independent study of SIDM in hydrodynamical simulations, focusing on low-mass dwarf galaxies. Their simulations employed bursty stellar feedback, which generated strong outflows and fluctuations in the gravitational potential, effectively creating dark matter cores even in the CDM model. As a result, the SIDM simulations with a cross-section of $\sigma/m = 2~\mathrm{cm}^2/\mathrm{g}$ showed little difference from their CDM counterparts in terms of central density structure. This outcome highlights the possibility that baryonic feedback can act as a systematic effect when interpreting small-scale discrepancies as evidence for SIDM. Therefore, testing both CDM and SIDM models under varied feedback prescriptions is essential for disentangling the roles of baryonic and dark matter properties.\\

More recently, the BAHAMAS (BAryons and HAloes of MAssive Systems) simulation suite \cite{McCarthy2017, McCarthy2018} has been extended to  include self-interacting dark matter models \cite{Robertson2019}. BAHAMAS employs large-volume cosmological hydrodynamical simulations with periodic boxes of $400~\mathrm{Mpc}/h$ per side, allowing for statistically robust predictions across a wide mass range, from galaxy groups to massive clusters. The simulations implement sub-grid physics models, which have been calibrated to reproduce key observables such as the galaxy stellar mass function, gas mass fractions, and X-ray luminosities of galaxy groups and clusters. 

Based on the BAHAMAS-SIDM simulations, additional signatures of dark matter self-interactions have been identified in galaxy clusters. Due to their higher velocity dispersions, clusters experience higher self-scattering rates than dwarf galaxies, resulting in more pronounced SIDM-induced cores. Robertson et al.\cite{Robertson2019} showed that the inclusion of baryons significantly modifies the density profiles of SIDM halos, with baryonic effects extending to larger radii compared to CDM counterparts. They also found that SIDM produces rounder dark matter halos in clusters (Figure~\ref{fig:Robertson2019}), with shape modifications extending out to scales comparable to the virial radius. However, the distributions of gas and stars remain unaffected, suggesting that gravitational lensing may provide the sensitive observational probe of SIDM-induced changes in halo shapes. 

Since SIDM tends to make the central regions of halos more spherical, it consequently produces smaller and more circular tangential critical curves in strong gravitational lensing—observable features that can be directly measured. In contrast, massive galaxy clusters in the CDM framework typically exhibit more elongated critical curves, reflecting their triaxial shapes formed through hierarchical structure formation. As a result, the critical curve of an SIDM halo of a given mass can resemble that of a less massive CDM halo, introducing a degeneracy with halo mass. However, when combined with independent mass measurements such as $M_{200}$, Robertson et al.~\cite{Robertson2019} showed that the size and axial ratio of the critical curve offer an approach to distinguish between dark matter models. Their work showed that incorporating baryonic physics into simulations yields more robust constraints on self-interacting dark matter, as baryons influence observable lensing signatures (and DM halo) both directly and indirectly.

In addition, Harvey et al.\cite{Harvey2017} used the BAHAMAS-SIDM simulations to investigate the wobbling of the Brightest Cluster Galaxy (BCG) within the center of galaxy clusters. Kim et al.~\cite{Kim2017} were the first to demonstrate that following the merger of two dark matter halos, the BCG can oscillate around the center of the remnant core for several Gyrs—persisting well after the cluster has relaxed. As a result, the detection of a miscentered BCG in a relaxed cluster can serve as indirect evidence for the presence of a central core in the dark matter density profile. In their analysis, Harvey et al.\cite{Harvey2017}  measured the radial offset between the large-scale dark matter halo and the BCG to estimate the wobble amplitude in the BAHAMAS simulations. However, they found this measurement to be sensitive to the simulation resolution. Using the high-resolution version of BAHAMAS, they constrained the wobble amplitude in the CDM scenario to be less than $A_w < 2\,\rm kpc$. Beyond theoretical predictions, Harvey et al. also analyzed observations of ten galaxy clusters from the Hubble Space Telescope and measured a wobble amplitude of $A_w = 11.82^{+7.3}_{-3.0}\,\rm kpc$. This result disfavors a zero-wobble amplitude at the $3\sigma$ confidence level, providing observational evidence consistent with the existence of a dark matter core in galaxy clusters.

Besides relaxed clusters, self-interacting dark matter (SIDM) also leave observable signatures in merging systems. In the standard collisionless CDM framework, collisionless dark matter and stars pass through each other nearly unaffected during cluster mergers, while the collisional intracluster gas experiences ram pressure and slows down, creating a separation between the gas and dark components. However, in SIDM models, dark matter particles can scatter with each other, resulting in a drag-like effect that causes dark matter to lag behind the stars—similar to the behavior of gas, leading to an additional offset between dark matter and stars that does not occur in the collisionless case \cite{Markevitch2004, Randall2008, Harvey2014, Harvey2015}.

To quantity this, a dimensionless quantity \textit{bulleticity} was introduced, defined as
\begin{equation}
\label{eq:bulleticity}
\beta_{\parallel} \equiv \frac{\delta_{\rm SD}}{\delta_{\rm SG}},
\end{equation}
where \( \delta_{\rm SD} \) is the spatial offset between the stellar and dark matter components, and \( \delta_{\rm SG} \) is the offset between stars and gas. This ratio captures the relative displacement of dark matter with respect to the gas and stars.

Sirks et al.\cite{Sirks2024} investigated the effects of SIDM on major mergers of galaxy clusters using the BAHAMAS-SIDM simulations. They found that the dependence of \( \beta_{\parallel} \) on the self-interaction cross-section \( \sigma/m \) follows an empirical relation predictions from Harvey et al.\cite{Harvey2015} :
\begin{equation}
\beta_{\parallel}(\sigma/m) = B \left( 1 - e^{-(\sigma/m)/A} \right).
\end{equation}
This relation captures the behavior of SIDM across different interaction regimes.
At low values of $\sigma/m$, the halo is optically thin, and the effective drag force increases approximately linearly with the cross-section. In contrast, at high $\sigma/m$, halo becomes optically thick, where dark matter particles interact frequently. This leads to a saturation in the drag force, and the resulting offsets between stars, gas, and dark matter components approach a constant value. 
%Observational constraints of $\sigma/m$ derived from these different features are presented in Section~\ref{sc:observe}.

In addition to the BAHAMAS-SIDM simulations, several recent hydrodynamical cosmological studies have investigated the role of self-interacting dark matter (SIDM) in structure formation. For example, the Dianoga SIDM simulations \cite{Ragagnin2024} examine galaxy clusters using both dark matter-only and full-physics zoom-in simulations, incorporating velocity- and angle-dependent cross-sections to compare rare (large-angle) and frequent (small-angle) scattering scenarios. They find that frequent scattering leads to stronger subhalo suppression and greater diversity in subhalo concentrations. Interestingly, when baryonic physics is included, the central densities in SIDM models become significantly higher than in their collisionless counterparts—a result that contrasts with findings from other simulations such as BAHAMAS \cite{Robertson2019}. This suggests that constraining the SIDM cross-section solely based on core density may be insufficient, highlighting the need for further studies on the complex interplay between SIDM and baryonic processes.

The AIDA-TNG project \cite{Despali2025} offers a suite of large-volume cosmological magnetohydrodynamic simulations that model galaxy formation under different dark matter frameworks. This includes warm dark matter and SIDM scenarios with both constant and velocity-dependent cross-sections. Spanning a wide halo mass range from $10^8$ to $4 \times 10^{14}~M_{\odot}$, AIDA-TNG allows for a systematic investigation of SIDM effects across a wide mass range of cosmic structures—from galaxies to galaxy clusters. These simulations also yield statistical predictions for key observables such as the halo mass function and the matter power spectrum, enabling direct comparison with large-scale structure surveys.

More recently, the DARKSKIES project \cite{Harvey2025} has extended SIDM studies through a suite of high-resolution hydrodynamical zoom-in simulations of galaxy clusters, exploring SIDM cross-sections spanning $\sigma/m = 0.01$--$0.2~{\rm cm^2,g^{-1}}$. By probing significantly smaller cross-sections than many previous simulation studies and resolving cluster cores at extremely high resolution, DARKSKIES is particularly well suited for investigating the subtle impact of self-interactions on the inner structure of galaxy clusters. The simulations provide predictions for halo density profiles, BCG offsets and wobbling, and other cluster-scale observables that can be directly compared with observations, thereby enabling increasingly stringent tests of SIDM models.

\section{Observational Constraints on the SIDM Cross-Section}
\label{sc:observe}
\begin{figure*}[htbp]
 \centering
 \includegraphics[width=1\textwidth,trim={0.5cm 0cm 0cm 0cm},clip]{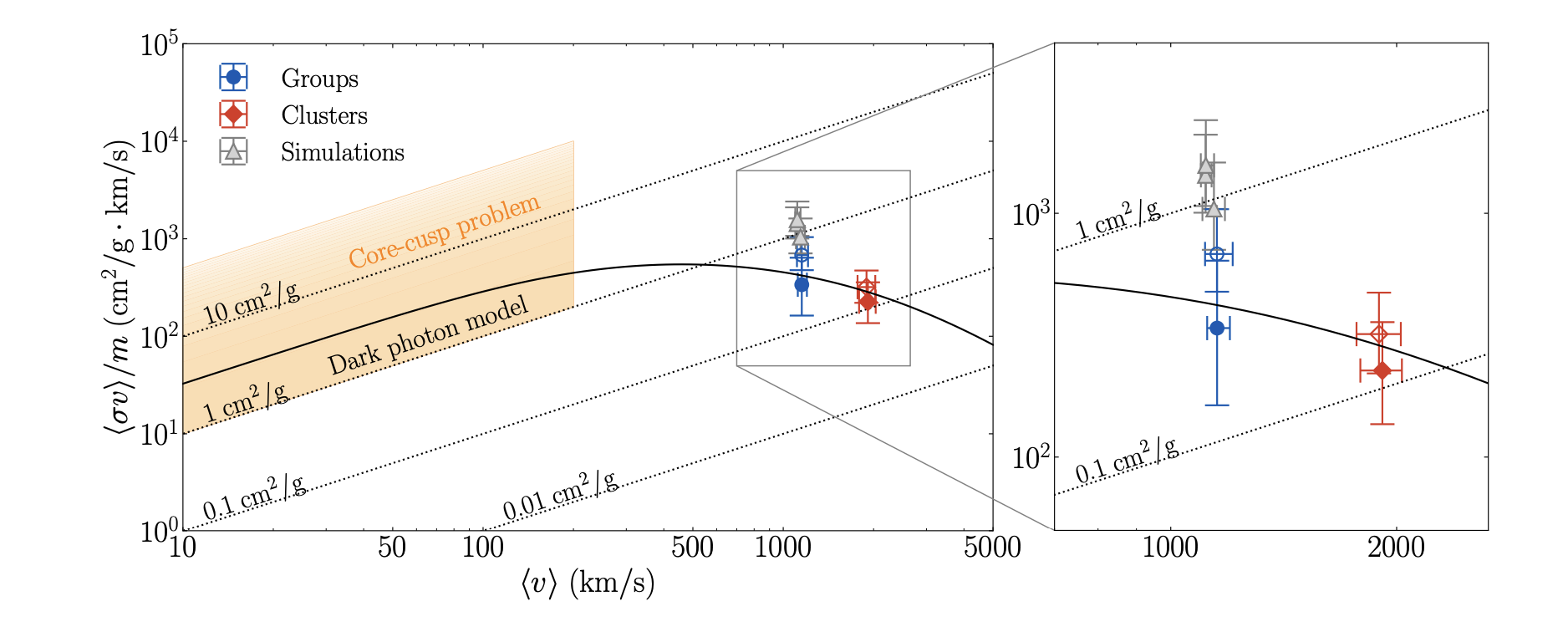}
\caption{Velocity dependence of self-interactions, shown as the mean velocity-weighted cross
section per unit mass $\sigma v/m$ versus mean scattering velocity $v$. The blue and red data points represent constraints from galaxy groups and galaxy clusters, respectively, based on the analysis by Sagunski et al. (2021). The shaded region indicates the preferred SIDM parameter space for resolving the core-cusp problem in dwarf galaxies. The solid curve corresponds to a particle physics model with 15 GeV dark matter interacting via an 11 MeV dark photon, which yields a velocity-dependent cross-section consistent with observational constraints across all halo mass scales. Reprinted from Sagunski et al. \cite{Sagunski2021}.
}
 
 \label{fig:Sagunski2021}
\end{figure*}

Self-interacting dark matter (SIDM) introduces modifications to the internal structure of dark matter halos, offering a compelling alternative to collisionless cold dark matter (CDM) for explaining certain small-scale astrophysical phenomena. Over the past decades, a wide range of observational studies—spanning from dwarf and low‑surface‑brightness (LSB) galaxies to galaxy groups and clusters\cite{Sagunski2021}  —have been employed to constrain the SIDM self-interaction cross-section, $\sigma/m$, by comparing inferred halo properties to theoretical predictions or numerical simulations that incorporate SIDM physics.
In galaxy clusters, where the typical relative velocity is around 1000 km/s, various probes—including strong gravitational lensing, X-ray emission, and cluster mergers—generally yield upper limits of
$\sigma/m \lesssim 0.1$--$1\,\mathrm{cm}^2\,\mathrm{g}^{-1}$, consistent with the collisionless scenario. In contrast, dwarf galaxies, which have much lower internal velocities (~30–100 km/s), often require significantly larger cross-sections in the range $\sigma/m\sim0.5-10\rm cm^2/g$ \cite{Dave2001,Valli2018,Ren2019} to explain their observed central cores.
Reconciling these disparate scales points to a velocity‑dependent cross‑section, which arises naturally if dark matter scatters through a light mediator. For example, Sagunski et al.~\cite{Sagunski2021} showed that a SIDM model with a $15\,\mathrm{GeV}$ dark matter particle and an $11\,\mathrm{MeV}$ dark-photon mediator naturally produces a velocity-dependent cross-section, yielding $\sigma/m\sim1\,\mathrm{cm^2\,g^{-1}}$ on galaxy scales and $\sim0.1\,\mathrm{cm^2\,g^{-1}}$ in galaxy clusters, consistent with observational constraints over a broad range of halo masses (Figure~\ref{fig:Sagunski2021}).

Since the  scattering rate depends on both the local dark matter density and the relative velocity (see Equation~\ref{eq:scatterrate}), galaxy clusters—with their high mass densities and large velocity dispersions—particularly sensitive laboratories for testing SIDM, allowing stringent constraints to be placed on the self-interaction cross-section.
In the following subsections, we review the principal observational probes used to constrain SIDM cross-section $\sigma/m$ in galaxy clusters.

\subsection{Observed halo density profile}
As discussed in Section~\ref{sec:density}, one of the most direct consequences of SIDM is its impact on the inner density profile of halos. This naturally suggests that measuring the density profile should provide a straightforward probe of dark matter self-interactions. However, in practice, this approach is challenging because the density profile is not a direct observable —it must be inferred through modeling based on indirect observables such as gravitational lensing, galaxy dynamics, or X-ray emission. Some inferences often rely on assumptions about the dynamical state or hydrostatic equilibrium, introducing uncertainties that can complicate the interpretation. Gravitational lensing provides a more direct probe of the matter distribution without assuming equilibrium: strong lensing is particularly sensitive to the inner regions most affected by SIDM \cite{Firmani2000, Firmani2001, Meneghetti2001}, while weak lensing typically constrains the mass distribution at larger radii. Meneghetti et al. \cite{Meneghetti2001} investigated the impact of dark matter self-interactions on strong lensing features in galaxy clusters, focusing on the formation of radial and giant tangential arcs. Their simulations showed that even a small number of collisions per particle can significantly alter the inner density structure of a cluster, reducing its ability to generate the highly elongated arcs observed in real systems and and suppressing the formation of radial arcs. Based on this, they concluded that the self-interaction cross-section must be small—specifically, $\sigma/m < 0.1,\mathrm{cm}^2/\mathrm{g}$—in order to remain consistent with the strong lensing features observed in clusters.

Recently, Andrade et al.~\cite{Andrade2022} performed strong lensing analyses of eight galaxy clusters to probe their inner density profiles over the radial range 10–150 kpc. By comparing the reconstructed mass distributions to predictions from SIDM halo models, they derived an upper limit on the self-interaction cross-section of $\sigma/m < 0.13,\mathrm{cm^2/g}$ at the 95$\%$ confidence level. 

Beyond gravitational lensing, multi-wavelength observations also provide independent constraints on SIDM. Eckert et al.~\cite{Eckert2022} analyzed the structural properties of 12 massive X-COP galaxy clusters using X-ray data from XMM-Newton and the Sunyaev–Zel’dovich (SZ) effect from the Planck all-sky survey. They measured the Einasto shape parameter of the dark matter distribution and found a mean value of $\alpha = 0.19 \pm 0.03$, consistent with expectations from the CDM framework. To interpret this in the context of SIDM, they adopted a power-law relation between the shape parameter and the cross-section per unit mass derived from the BAHAMAS-SIDM simulations:

\begin{equation}
\alpha_{\mathrm{Einasto}} = \alpha_0 + \alpha_1 \left( \frac{\sigma/m}{1\mathrm{cm}^2/\mathrm{g}} \right)^\gamma.
\end{equation}

Using this relation, they converted the observed constraint on $\alpha$ into a 95\% confidence upper limit on the SIDM cross-section of $\sigma/m < 0.19,\mathrm{cm^2/g}$ corresponding to typical cluster-scale collision velocities of $v \sim 1000\mathrm{km/s}$.

Adhikari et al.~\cite{Adhikari2025} recently used weak-lensing measurements from the Dark Energy Survey (DES) Year 3 data \cite{Sevilla-Noarbe} of around $\sim$1000 galaxy clusters identified via the Sunyaev–Zeldovich (SZ) effect in the Advanced Atacama Cosmology Telescope (AdvACT) survey to test predictions from SIDM models. By comparing the stacked weak-lensing profiles with simulated SIDM halos, they constrained the isotropic and elastic self-interaction cross-section to $\sigma/m < 1\mathrm{cm^2/g}$. This upper limit is weaker than those derived from strong lensing analyses, likely due to the fact that weak lensing primarily probes the outer halo regions. Measurements start from $r \sim 0.2h^{-1}\mathrm{Mpc}$ in their studies, where the impact of dark matter self-interactions on the density profile is expected to be substantially weaker.

In addition, stellar kinematics of the brightest cluster galaxy (BCG) offer a valuable constraint on the central mass distribution, helping to disentangle the contribution from baryons and refine the slope of the dark matter density profile in the core \cite{Newman2013,He2020}. Newman et al. \cite{Newman2013} performed a joint analysis of stellar kinematics of the brightest cluster galaxies (BCGs) and gravitational lensing data for seven massive galaxy clusters. They found that, while the total mass density profiles are broadly consistent with NFW predictions at large radii, the inner regions (r $<$ 30 kpc), comparable to the effective radii of the BCGs, can be equally well described by cored NFW profiles with core radii of approximately 10–20 kpc. However, interpreting such cores as evidence for SIDM is complicated by the dominant contribution of baryons in the central regions of galaxy clusters, which can significantly modify the dark matter distribution through gravitational coupling. Multi-probe analyses require high-quality lensing and stellar kinematic data, which are rarely available for the same sample of dynamically relaxed clusters. Combining all probes also demands a careful joint modeling framework with a self-consistent treatment of systematics and cluster dynamical states. Therefore, while multi-probe analyses have advanced, comprehensive studies that jointly model stellar kinematics and lensing data within the SIDM framework remain limited.

\subsection{Halo Shape Measurement}
Dark matter self-interaction leads to thermalization of the inner halo regions, where the particle collision rate is highest. This thermalization tends to isotropize the velocity dispersion and redistribute energy, thereby making the central regions of halos more spherical than predicted by the collisionless cold dark matter (CDM) paradigm. Both the SIDM-only simulations by Brinckmann et al.\cite{Brinckmann2018} and the hydrodynamical simulations by Robertson et al.\cite{Robertson2019} demonstrate that halo shapes are affected by dark matter self-interactions on larger spatial scales than their density profiles, which suggests that halo shape may provide a more robust and less baryon-sensitive probe of dark matter self-interactions compared to inner density structure.
Therefore, one potential method to test for these differences is through measurements of the shapes of dark matter halos. In particular, the projected ellipticity of a galaxy cluster’s mass distribution can be probed via gravitational lensing. Miralda-Escudé \cite{Miralda-Escude} was the first to use ellipticity estimates at 70 kpc from the strong-lensing analysis of the galaxy cluster MS2137-23  \cite{Mellier1993,Miralda-Escude1995} to constrain the dark matter self-interaction cross-section. By assuming that the dark matter halo would appear spherical if the  scattering rate exceeded the Hubble rate, i.e., $\Gamma \gtrsim H_0$, Miralda-Escudé \cite{Miralda-Escude} obtained a strong constraints on $\sigma/m \lesssim 0.02\,\mathrm{cm^2/g}$. However, subsequent studies incorporating N-body simulations showed that this constraint was likely overestimated, as the relationship between SIDM-induced sphericity and projected ellipticity is more complex than initially assumed. 

Peter et al. \cite{Peter2013} used cosmological simulations to study  dark matter halo shape under self-interacting dark matter (SIDM) models with cross-sections of $\sigma/m=0.03, 0.1, 3\rm\,cm^2/g$.  They found that even 1 $\rm cm^2/g$ can still satisfy strong lensing results for MS2137-23. This mainly is because one scattering event per particle is not enough to remove all triaxiality
from a DM halo based on simulation results. Moreover, because lensing observables are sensitive to the projected surface mass density, the measured ellipticity at small projected radii still receives contributions from outer halo regions—where the self-interaction rate is low and the shape may remain elliptical—making it difficult to isolate the core shape changes induced by self-interactions \cite{Peter2013,TulinYu}. They also found that the intrinsic scatter in the halo axis ratio (c/a) is typically around 10–20\%, driven by variations in the mass assembly history and environment of each system. As a result, robust constraints on SIDM based on halo shapes should be derived from statistical samples of clusters rather than relying on an individual system. Peter et al. therefore conducted the  shape-based constraints based on five clusters from the Local Cluster Substructure Survey (LoCuSS)\cite{Richard2009} and compare with the five most massive simulated halos with similar range of velocity dispersion $\sigma_0$ and core-radius $r_{\rm core}$. They found that the observed ellipticity distribution of the LoCuSS clusters was more consistent with SIDM models having
$\sigma/m=0.1\,\mathrm{cm^2\,g^{-1}}$
than with models having
$\sigma/m=1\,\mathrm{cm^2\,g^{-1}}$.

Besides strong lensing, weak gravitational lensing is also powerful to reconstruct the projected mass distribution over large scales.  Robertson et al.~\cite{Robertson2023} investigated whether cluster shapes inferred from weak gravitational lensing measurements could serve as an effective probe of the self-interacting nature of dark matter, using the BAHAMAS-SIDM simulations. However, they found that the distributions of weak-lensing-inferred cluster shapes are nearly indistinguishable between the CDM and SIDM scenarios. This limitation arises primarily due to projection effects inherent in lensing observables, as well as the fact that weak lensing is most sensitive to scales near the virial radius—where the structural differences between SIDM and CDM halos are minimal. Their results suggest that weak-lensing measurements alone are unlikely to provide competitive constraints on SIDM unless combined with probes that are sensitive to the inner halo structure.

Taken together, current halo-shape measurements provide evidence that SIDM can modify the morphology of cluster-scale halos, but translating observed ellipticities into robust constraints on the self-interaction cross-section remains challenging. Projection effects, intrinsic halo-to-halo scatter, and the contribution of outer halo regions all weaken the connection between observed shapes and the underlying scattering rate. Consequently, present shape-based studies generally favor
$\sigma/m \lesssim 0.1$--$1\,\mathrm{cm^2\,g^{-1}}$,
while highlighting the need for large statistical samples and realistic simulations to fully exploit halo shapes as a probe of SIDM.

\subsection{Cluster Mergers}
Merging systems of galaxy clusters, particularly nearly head-on collisions, provide another promising environment to test for dark matter self-interactions, as discussed in Section 4.1.3. In the SIDM scenario, collisions between DM particles during cluster mergers can lead to a collisional drag effect. This drag causes a spatial offset between the distribution of dark matter and the collisionless member galaxies. Several massive merging clusters have been studied extensively through multi-wavelength observations, including the Bullet Cluster \cite{Markevitch,Randall}.
For a summary of SIDM constraints derived from individual major merging clusters, we refer readers to Table II of Tulin and Yu \cite{TulinYu}. These individual systems yield constraints on the self-interaction cross-section in the range of $\sigma/m \lesssim 1.25$–$4\mathrm{cm}^2/\mathrm{g}$.
Mass distribution in such systems can also be reconstructed using weak lensing. Harvey et al.\cite{Harvey2015} performed a stacked weak-lensing analysis of 30 merging systems—both major and minor—to measure the bulleticity (defined in Equation~\ref{eq:bulleticity}). They found $\beta_{\parallel} = -0.04 \pm 0.07$, consistent with collisionless CDM, placing an upper limit on the self-interaction cross-section of $\sigma/m \lesssim 0.47\,\rm cm^2/g$. However, Wittman et al. \cite{Wittman} point out that some of the offset between stellar and dark matter components in the Harvey et al. sample was derived using only single-band imaging. These estimates are inconsistent with more detailed studies that employed multiband and strong lensing-based centroid measurements, which provide more accurate position measurements. After considering more accurate offset measurements, they revised the constraint to $\sigma/m \lesssim 2.0\,\mathrm{cm}^2/\mathrm{g}$.

While the studies above rely on measuring offsets between galaxies and dark matter in merging clusters, recent work has explored alternative observables that are less sensitive to uncertainties in merger geometry and dynamical phase. Jee et al. \cite{Jee2026} introduced a new approach using the ratio between the shock-to-shock separation traced by radio relics and the halo-to-halo separation measured from weak lensing. Since dark matter self-interactions can reduce the halo-to-halo separation through collisional drag while having little effect on the propagation of merger shocks, this observable provides a sensitive probe of the SIDM cross-section. Applying this technique to a sample of eleven cluster mergers hosting symmetric double radio relics, they obtained a 68\% upper limit of $\sigma/m < 0.22\,{\rm cm^2\,g^{-1}}$, while explicitly marginalizing over uncertainties associated with the merger geometry and dynamical state. This work highlights the potential of multi-wavelength observations, combining radio and gravitational-lensing measurements, as a complementary approach to constraining dark matter self-interactions in galaxy clusters.

In addition to signatures arising during cluster mergers, SIDM can also leave observable imprints in the post-merger evolution of galaxy clusters. Following a merger, BCGs in SIDM halos can remain on long-lived oscillatory orbits, leading to a offset from the halo center of mass—even in relaxed systems \cite{Kim2017, Harvey2017} (see Section 4.1.3). Harvey et al.~\cite{Harvey2019} analyzed 10 strong lensing clusters and, by comparing the observed BCG–halo offsets with predictions from BAHAMAS SIDM simulations, placed a constraint on the self-interaction cross-section of $\sigma/m\lesssim0.39\,\mathrm{cm}^2/\mathrm{g}$. More recently, Cross et al.\cite{Cross2024} examined 23 relaxed galaxy clusters from the Dark Energy Survey and the Sloan Digital Sky Survey, focusing on the offset between the X-ray peak and the central galaxy. They found a non-zero median offset of $\mu=6.0^{+1.4}_{-1.5}\rm \, kpc$, which is qualitatively consistent with the long-lived oscillations predicted in some SIDM models. However, further simulation work is required to interpret these offsets quantitatively and convert them into constraints on $\sigma/m$.

In summary, cluster mergers provide one of the most direct astrophysical tests of dark matter self-interactions, as they probe the collisional behavior of dark matter during and after high-velocity encounters. While constraints derived from individual merging systems typically lie in the range
$\sigma/m \lesssim 1$--$4\,\mathrm{cm^2\,g^{-1}}$,
recent statistical analyses and newly developed multi-wavelength observables have improved these limits to
$\sigma/m \lesssim 0.2$--$0.5\,\mathrm{cm^2\,g^{-1}}$.

\subsection{Galaxy-Galaxy Strong-Lensing Excess}
Beyond the well-known small-scale challenges discussed in Section~\ref{sc:small-scale}, 
Meneghetti et al.~\cite{Meneghetti2020} reported a significant excess of small-scale strong gravitational lenses in galaxy clusters. In particular, cluster substructures were found to be more efficient lenses than predicted by standard $\Lambda$CDM simulations, with galaxy-galaxy strong-lensing (GGSL) probabilities exceeding theoretical expectations by more than an order of magnitude.

To assess whether this discrepancy could be attributed to baryonic physics or simulation systematics. Meneghetti et al.~\cite{Meneghetti2023} analyzed a broader suite of hydrodynamical simulations using different galaxy formation models. Even when employing the GIZMO-SIMBA simulations—which produce denser stellar cores and enhance the  galaxy-galaxy strong lensing efficiencies by a factor of $\sim 3$—the predicted lensing frequencies remained lower than observed, indicating that current $\Lambda$CDM-based cosmological hydrodynamical simulations cannot fully account for the excess.

Tokayer et al.~\cite{Tokayer2024} further explored the role of baryonic processes by incorporating adiabatic contraction into CDM subhalos to model the impact of baryons. They found that the inclusion of baryons can steepen the inner slope of density profile $\rho\propto r^{\gamma}$, but only up to $\gamma < 2.0$, which remains insufficient to account for the high strong-lensing efficiencies seen in observations.

These findings suggest that modifications to the nature of dark matter may be necessary. In particular, self-interacting dark matter (SIDM) models offer a compelling alternative. Yang and Yu~\cite{YangYu} proposed that subhalos in SIDM cluster environments, particularly those hosting early-type galaxies, could undergo gravothermal core collapse, leading to steeper central density profiles and enhanced lensing cross-sections. Dutra et al.~\cite{Dutra2025} further explored this by demonstrating that increasing the inner density slope from a CDM-like $\rho \propto r^{\gamma}$ with $\gamma \sim 1$ to values exceeding $\gamma > 2.5$ can substantially alleviate the GGSL discrepancy. This steep inner slope can not be explained by the inclusion of baryons in CDM halos \cite{Tokayer2024}. Therefore, as emphasized by Dutra et al.\cite{Dutra2025}, high-resolution hydrodynamical simulations of SIDM on galaxy cluster scales, with sufficient resolution to capture the collapse of subhalo cores, will be crucial to assess whether SIDM can offer a solution to the GGSL excess.

\subsection{Halo RAR}
A more recent approach to constraining dark matter self-interactions is through the radial acceleration relation (RAR) of galaxy clusters.
The RAR is a tight empirical correlation between the total gravitational acceleration, $g_{\rm tot} = GM_{\rm tot}(<r)/r^2$, and the baryonic component, $g_{\rm bar} = GM_{\rm bar}(<r)/r^2$. This relation was first established by McGaugh et al.~\cite{McGaugh2016}, who analyzed the rotation curves of 153 spiral galaxies and found that these two independently measured quantities, $g_{\rm tot}$ and $g_{\rm bar}$, follow a relation: 

\begin{equation}
\frac{g_{\rm tot}}{g_{\rm bar}} = \frac{M_{\rm tot}}{M_{\rm bar}} = \frac{1}{1 - e^{-\sqrt{g_{\rm bar}/g_\dagger}}},
\label{eq:McGaugh}
\end{equation}
where $g_{\dagger} = (1.20 \pm 0.24) \times 10^{-10}\mathrm{m\,s^{-2}}$ is a characteristic acceleration scale.
Subsequent studies have confirmed the existence of this relation across diverse galaxy samples and shown that it is independent of other galaxy properties such as  baryonic mass, morphological types or gas fraction \cite{Lelli, Rong2018, Chae, Brouwer}. On the theoretical side, hydrodynamical simulations within the $\Lambda$CDM framework have successfully reproduced the observed RAR of galaxies \cite{Keller, Ludlow, Garaldi, Paranjape}.

While originally discovered as a tight relation in galaxies, recent studies have extended the analysis to cluster-scale objects \cite{Tian, Pradyumna, Eckert2022a}. The first cluster-scale study was performed by Tian et al. \cite{Tian} which analyzed 20 high-mass galaxy clusters from CLASH program. In their work, the total mass profiles were derived through a joint analysis of strong and weak gravitational lensing data \cite{Umetsu2016}, while the baryonic mass was estimated by X-ray gas mass measurements and stellar mass estimates \cite{Donahue2014}. The resulting RAR for galaxy clusters was found to deviate significantly from the relation observed in galaxies: the characteristic acceleration scale, $g_{\dagger}$, was measured to be approximately an order of magnitude higher than that obtained from galaxy-scale systems. This finding suggests that the RAR is not a universal relation that holds across all mass scales.

It is also worth noting that the tightness of the galaxy RAR has often been cited as empirical support for Modified Newtonian Dynamics (MOND) \cite{Milgrom}, which can reproduce the observed relation without invoking dark matter. However, the departure of the cluster-scale RAR from the galaxy RAR poses a challenge for simple MOND interpretations based on a universal acceleration scale. These cluster-scale results indicate that the RAR may instead emerge from the complex interplay of baryonic physics and dark matter in structure formation, rather than reflecting a fundamental modification of gravity.

Therefore, the RAR offers a potentially powerful probe of dark matter physics. Tam et al.~\cite{Tam2023} investigated the RAR of galaxy clusters using the BAHAMAS-SIDM simulations and found a strong dependence of the RAR slope on the self-interaction cross-section $\sigma/m$. Specifically, for higher values of $\sigma/m$, the RAR becomes shallower, reflecting the reduced total gravitational acceleration caused by the formation of SIDM-induced cores. This deviation is most pronounced in the central regions of clusters, particularly near the brightest cluster galaxy (BCG). Notably, they also found that the differences in the RAR between CDM and SIDM models are more significant than the corresponding cusp–core differences in the density profiles at radii beyond $100\,\rm kpc$. These results highlight the potential of the RAR, when constrained by joint strong and weak lensing observations, as a sensitive probe to discriminate between SIDM and CDM scenarios.

\begin{figure*}[htbp]
 \centering
 \includegraphics[width=0.9\textwidth,trim={0.5cm 0cm 0cm 0cm},clip]{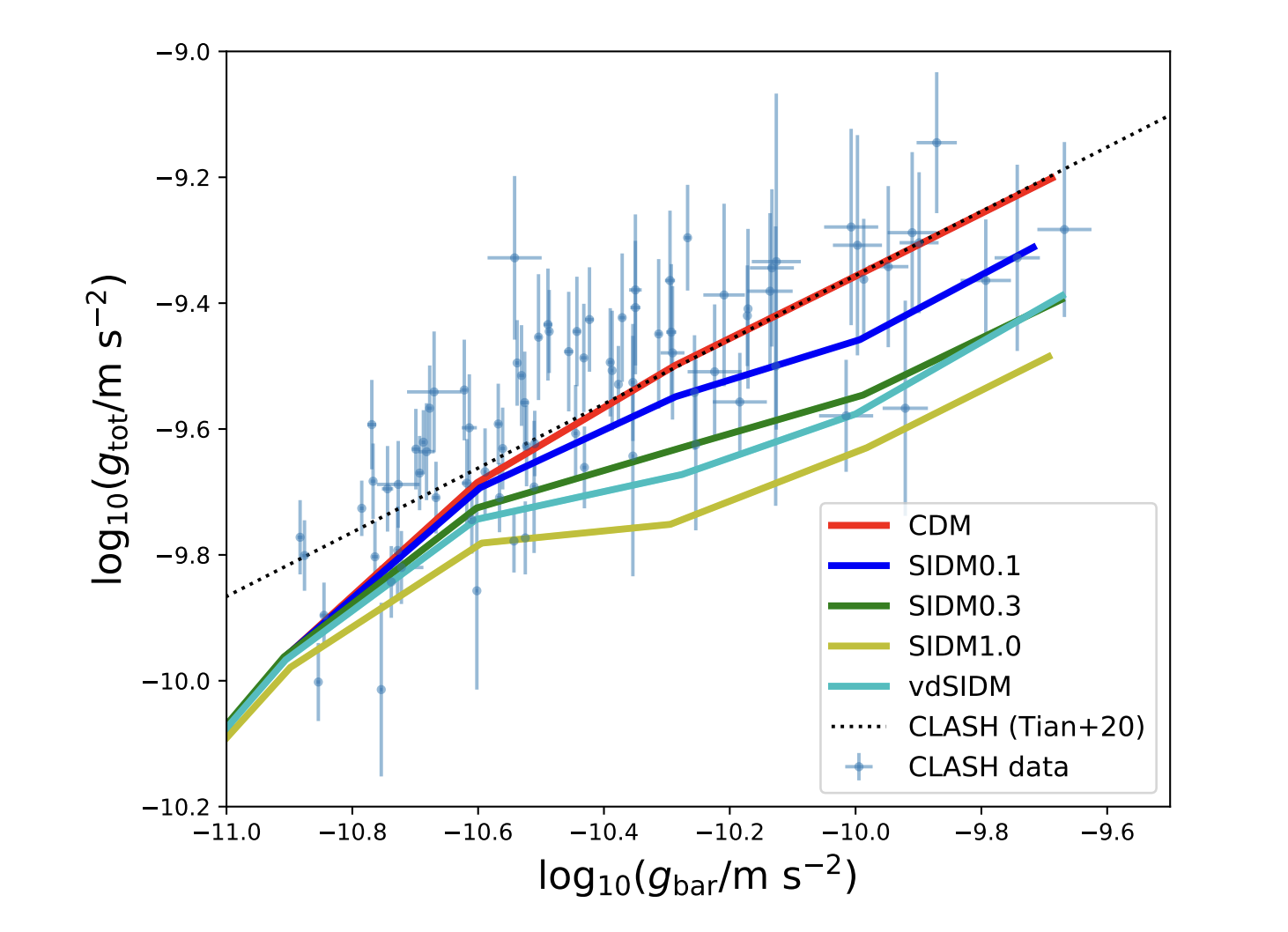}
\caption{Comparison between the observed RAR from CLASH clusters (Tian et al.~\cite{Tian}) and the theoretical predictions from the BAHAMAS simulations under five different dark matter models. The blue points with error bars represent the CLASH measurements of cluster centripetal accelerations. Solid lines indicate the predicted RAR from BAHAMAS simulations for each dark matter model.  
 Reprinted from Tam et al.~\cite{Tam2023}.
}
 
 \label{fig:RAR}
\end{figure*}

They quantitatively compared the theoretical predictions from the BAHAMAS-SIDM simulations with the observational RAR data from the CLASH clusters analyzed by Tian et al.\cite{Tian}. As shown in Figure~\ref{fig:RAR}, the observed RAR is consistent with the CDM scenario, including the high-acceleration regime associated with the brightest cluster galaxies (BCGs). Based on their statistical analysis, even when conservatively excluding the BCG regions, SIDM models with $\sigma/m = 0.3~\mathrm{cm^2g^{-1}}$ were found to be at the $3.8\sigma$ level relative to CDM. This study highlights the power of the cluster-scale RAR as a probe for testing the collisionless nature of dark matter. In particular, because SIDM effects are most prominent in the inner regions of clusters, future observations that provide more precise measurements of the dynamical accelerations in BCGs via stellar kinematics \cite{Tian2024} will enable even more stringent constraints on the self-interaction cross-section $\sigma/m$.

\label{sec:cluster_SIDM}
\section{Summary and Future Aspects}
In this review, we examined the success of CDM model in explaining large-scale cosmic structures and evolution of large-scale cosmic structures, as well as the challenges it faces on small scales. As an alternative, we explored the SIDM model, which introduces dark-matter self-interactions that can modify the internal halo structures. We reviewed the theoretical foundations of SIDM, the predictions from numerical simulations, and current observational constraints derived primarily from galaxy cluster-scale measurements.

Over the past two decades, SIDM has evolved from a proposed solution to the small-scale challenges of CDM into a mature framework that can be tested using increasingly precise observations across a wide range of astrophysical systems. However, the interpretation of many astrophysical signatures remains complicated by baryonic processes, which can produce effects similar to those expected from dark matter self-interactions. Recent work by Harvey \cite{Harvey2024} applied a deep-learning method to distinguish between the impact of dark matter self-interactions and that of astrophysical feedback, using simulated weak-lensing mass maps. This opens up a powerful tool for interpreting observational data and breaking degeneracies that limited our ability to constrain the fundamental properties of dark matter.

In the future, progress in understanding the nature of dark matter will rely on both improved observational data and sophisticated theoretical modeling. Upcoming large-scale surveys such as the LSST, {\it Euclid}, and {\it Roman Space Telescope} will provide high-quality data on the distribution and evolution of cosmic structures across cosmic time, enabling tighter constraints on dark matter models and helping to test different dark matter scenarios. Achieving this goal will also require advanced hydrodynamical simulations that incorporate both baryonic feedback and self-interacting dark matter physics, enabling the complex interplay between these processes to be disentangled.

Besides astronomical constraints, progress in understanding dark matter also depends on insights from particle physics. A CDM candidate is often hypothesized to be a weakly interacting massive particle (WIMP), motivated by extensions of the Standard Model. Numerous experimental efforts have been undertaken to detect such particles—either directly in underground detectors, via production in high-energy colliders, or indirectly through the detection of annihilation or decay products. However, despite decades of intensive searches, no conclusive detection has yet been achieved. On the other hand, various particle physics models have been proposed to describe self-interacting dark matter (SIDM), introducing different dark-sector interactions that govern the self-scattering interactions. Determining key properties—such as the dark matter particle mass, self-interaction cross section, and the characteristics of the mediator—remains a key research topic. Ultimately, uncovering the nature of dark matter will require a combined effort across cosmology, astrophysics, numerical simulations, and particle physics. Continued synergy between these fields will be essential for revealing the fundamental properties of dark matter and its role in the formation of cosmic structures.

\section*{Acknowledgments}
The author would like to thank Prof. Wei-Tou Ni for the invitation to contribute this review article to the commemorative volume \textit{110 Years of General Relativity}. The author is grateful to David Harvey, Andrew Robertson, Keiichi Umetsu, and other colleagues for helpful discussions on self-interacting dark matter and galaxy cluster studies. The author also thanks the anonymous referee for constructive comments and suggestions that improved the quality and clarity of the manuscript.

%\begin{thebibliography}{000} %for 3 digits
%\begin{thebibliography}{00}  %for 2 digits

\end{document}